\documentclass[prb,twocolumn,preprintnumbers,amsmath,amssymb,superscriptaddress,longbibliography,nofootinbib]{revtex4}

\usepackage{algorithm}
\usepackage{mathtools}
\usepackage{algpseudocode}

\usepackage{color}
\usepackage{graphicx}% Incl fgfrude figure files
\usepackage{dcolumn}% Align table columns on decimal point
\usepackage{bm}% bold math
\usepackage{hyperref}
\usepackage{enumitem}
\usepackage{balance}
\usepackage{comment}
\usepackage{cleveref}
\usepackage{amsthm}

\newtheorem{lemma}{Lemma}

\newcommand{\tr}{{\rm tr}}

\makeatletter
\newsavebox{\@brx}
\newcommand{\llangle}[1][]{\savebox{\@brx}{\(\m@th{#1\langle}\)}%
  \mathopen{\copy\@brx\kern-0.5\wd\@brx\usebox{\@brx}}}
\newcommand{\rrangle}[1][]{\savebox{\@brx}{\(\m@th{#1\rangle}\)}%
  \mathclose{\copy\@brx\kern-0.5\wd\@brx\usebox{\@brx}}}
\makeatother

\begin{document}
%\title{Operator Evolution Dimension in polynomially sized equivalence classes of Pauli string}
%\title{Polynomially Restricted Operator Growth in equivalence classes of Pauli strings}
\title{Polynomially restricted operator growth in dynamically integrable models}

\author{Igor Ermakov}
\affiliation{Russian Quantum Center, Skolkovo, Moscow 121205, Russia}
\affiliation{Department of Mathematical Methods for Quantum Technologies, Steklov Mathematical Institute of Russian Academy of Sciences, 8 Gubkina St., Moscow 119991, Russia.}

\author{Tim Byrnes}
 \email{tim.byrnes@nyu.edu}
\affiliation{New York University Shanghai, NYU-ECNU Institute of Physics at NYU Shanghai, 567 West Yangsi Road, Shanghai, 200126, China.}
\affiliation{State Key Laboratory of Precision Spectroscopy, School of Physical and Material Sciences, East China Normal University, Shanghai 200062, China}
\affiliation{Center for Quantum and Topological Systems (CQTS), NYUAD Research Institute, New York University Abu Dhabi, UAE.}
\affiliation{Department of Physics, New York University, New York, NY 10003, USA}

\author{Oleg Lychkovskiy}
\email{lychkovskiy@gmail.com}
\affiliation{Skolkovo Institute of Science and Technology, Bolshoy Boulevard 30, bld. 1, Moscow 121205, Russia.}
\affiliation{Russian Quantum Center, Skolkovo, Moscow 121205, Russia}
\affiliation{Department of Mathematical Methods for Quantum Technologies, Steklov Mathematical Institute of Russian Academy of Sciences, 8 Gubkina St., Moscow 119991, Russia.}

\begin{abstract}
We provide a framework to determine the upper bound for the complexity of computing the exact Heisenberg representation of a given operator with respect to a Hamiltonian. Working in Heisenberg picture, we show that each Hamiltonian defines an equivalence relation, causing the operator space to be partitioned into equivalence classes. Any operator within a specific class never leaves it during the evolution. We provide a method to determine the dimension of the equivalence classes and evaluate it for various models, such as the $XY$ chain and Kitaev model on trees. 
For classes with a dimensionality that is not excessively large, 
one can use our framework to compute exact Heisenberg representation of any operator within this class. Our findings reveal that the complexity of operator evolution in these models grows from the edge to the bulk, which is physically manifested as suppressed relaxation of qubits near the boundary. Our methods are used to reveal several new cases of simulable quantum dynamics, such as $XY$ evolution augmented with certain non-Clifford single-qubit gates. We also introduce a $XY$-$ZZ$ model where the dimension of an equivalence class, while exponential in the system size, grows considerably slower than could be naively anticipated,  which entails practical computational advantage when simulating finite systems. Furthermore, we demonstrate how to apply our method to time-dependent Hamiltonians and dissipative systems.
\end{abstract}

\date{\today}

%\pacs{}
\maketitle

\section{Introduction}
Solving time dynamics of quantum many-body problems is typically an extremely difficult task.  In a brute force computation of such systems, one requires solving a coupled differential equations with an exponential number of variables with respect to the system size. In the Heisenberg formulation, an initially local observable evolves into an exponentially complex sum of non-local operators \cite{swingle2018unscrambling,parker2019universal,li2017measuring,von2018operator,nahum2018operator,khemani2018operator}. This proliferation of operators in the Heisenberg picture, called operator growth, has been explored recently in numerous closed \cite{von2018operator,khemani2018operator,swingle2018unscrambling,parker2019universal,maccormack2021operator,ballar2022krylov,garcia2022out,zhou2023operator,uskov2024quantum} and open quantum systems \cite{zhang2019information,yoshida2019disentangling,touil2021information,Schuster_2023_Operator,Rakovszky_2022_Dissipation-assisted,de2024stochastic,ermakov2024unified}.  Understanding operator complexity is crucial for determining the boundary between solvable and unsolvable problems in quantum dynamics, distinguishing between reachable and non-reachable quantum states, and understanding how information propagates through quantum systems. In a quantum computing context, the simulability of quantum circuits is a crucial question in the context of quantum advantage and other applications \cite{arute2019quantum,aaronson2004improved,boixo2018characterizing,smith2019simulating,wu2021strong,madsen2022quantum,xu2018emulating}.  

In some special cases, the space where the operator evolution occurs may not necessarily cover the full operator space.  For instance, in free-fermionic spin chains, the $z$-spin projection operator under Heisenberg evolution only scales polynomially with the system size \cite{valiant2001quantum,valiant2002expressiveness}. This restricted evolution explains the solvability of various dynamic \cite{Prosen_1998_New,Zunkovic_2014_Closed,Foss-Feig_2017_Solvable,Shibata_2019_Dissipative_Kitaev_chain,Dolgirev_2020_Non-Gaussian,Essler_2020_Integrability,lychkovskiy2021closed,Budini_2021_Solvable,gamayun2021nonequilibrium,guo2018analytical,horstmann2013noise,gamayun2022out,Teretenkov_2024} and transport \cite{vznidarivc2010exact,vznidarivc2013transport,ermakov2024effect,vznidarivc2010exact,ghosh2023relaxation,Ferreira_2024} problems, as well as the classical simulability of matchgate circuits \cite{valiant2001quantum,valiant2002expressiveness,jozsa2008matchgates,jozsa2013classical,brod2016efficient}. Moreover, different local operators within the same system can evolve in non-overlapping subspaces with varying dimensions and growth patterns. Examples include the $z$- and $x$-spin projections in the $XX$ model \cite{parker2019universal}, or edge modes decoupled from the bulk \cite{kitaev2001unpaired,fendley2016strong,fendley2012parafermionic,mi2022noise} discovered in various models. The dimensionality of this subspace, which we call the Operator Evolution Dimension (OED), provides a natural upper bound for the complexity of simulating the corresponding quantum dynamics problem. We refer to the models where the OED is polynomial in system size as dynamically integrable.

In this paper, we introduce a framework to classify the OEDs of different local operators, which allows for a powerful way of understanding the difference between simulable and non-simulable quantum dynamics problems. Given a spin-$1/2$ Hamiltonian, we show that one may define a equivalence relation on the set of Pauli strings, such that the span of each equivalence class is closed with respect to commutation with the Hamiltonian. We then study how specific Hamiltonians segregate the space of Pauli strings into these equivalence classes and compute the corresponding OEDs for operators within these classes. In the case of $XY$ dynamics we derive exact expressions for OEDs as integer-valued polynomials. Such knowledge of OEDs gives insight into the complexity of quantum dynamics of the system, which we illustrate on several models, including some which are non-free fermion models. In particular, we fully characterize OEDs of all local operators in $XY$ Hamiltonians.  To our knowledge, such a characterization was only available before for a very limited set of local operators that are quadratic in fermionic representation.

\section{Equivalence classes of Pauli strings} In the operator space of $L$ spin-1/2 particles, we define a Pauli string as $P=S_1\otimes \dots \otimes S_L$, where $S_i\in\{I_i,X_i,Y_i,Z_i\}$. Here $X_i,Y_i,Z_i$ are Pauli matrices acting on $i$th site and $I_i$ is the identity matrix. There are $4^L$ different Pauli strings in total that constitute a complete operator basis $\mathcal{P}=\{P_n\}^{4^L}_{n=1}$. 

Consider an operator $A\in\mathcal{P}$, such that $A\equiv P_n$. The aim in a time dynamical problem is to find the Heisenberg representation of this operator $A(t)$, given by following expansion:
\begin{align}
\label{eq:Bd_expanded}
A(t)\equiv \sum\limits^{D}_{m=1}f_{m}(t)P_m,
\end{align}
where $f_{m}(t)$ are time-dependent functions that should be obtained from the solutions of Heisenberg equations and $D$ is the OED of the operator $A$. The evolution is governed by the Hamiltonian $H$ that can be represented as a sum of $M$ Pauli strings, 
\begin{align}
\label{ham_lin}
H=\sum\limits_{n=1}^M h_nH_n, 
\end{align}
where $H_n\in\mathcal{P}$ are Pauli strings (referred to as Hamiltonian strings) and $h_n\neq 0$ are real numbers. 

Given the set of Hamiltonian strings $\mathcal{H} = \{ H_n \}_{n=1}^M $, we can define an equivalence relation among Pauli strings (denoted by $\sim$) within the set $\mathcal{P}$. Specifically, we consider strings $P_n$ and $P_m$ as equivalent if they coincide or if one can be derived from another after $Q$ rounds of commutation with Hamiltonian strings from $\mathcal{H}$. More specifically, there exists $\{H_{k_1},\dots,H_{k_Q}\}\in\mathcal{H}$ and a complex number $a$ such that:
\begin{align}
\label{pauli_binary_rel}
P_n=a[H_{k_Q},\dots,[H_{k_2},[H_{k_1},P_m]]\dots ].
\end{align} 
This binary relation satisfies the axioms of equivalence relations. This implies reflexivity $P_n\sim P_n$, symmetry\footnote{This follows from $[H_k,[H_k,P]]=4 P$ for any two non-commuting Pauli strings $H_k$ and $P$.} $P_n\sim P_m \iff P_m\sim P_n$, and transitivity: if $P_n\sim P_m$ and $P_m\sim P_l$ then $P_n\sim P_l$.   

The above equivalence relation results in a partition of $\mathcal{P}$ into $K$ disjoint equivalence classes $\mathcal{P}=\mathcal{A}^1\bigcup\dots \bigcup\mathcal{A}^K$. We will employ the notation $\mathcal{A}[A]$ for equivalence classes, where $A$ is some particular Pauli string from this class, the corresponding OED is denoted as $D[A]$. During the evolution the operator $A$ never leaves its class $\mathcal{A}$. Note that to define an equivalence class, it is sufficient to specify just one operator from that class. 

\begin{figure}
  \includegraphics[width=\columnwidth]{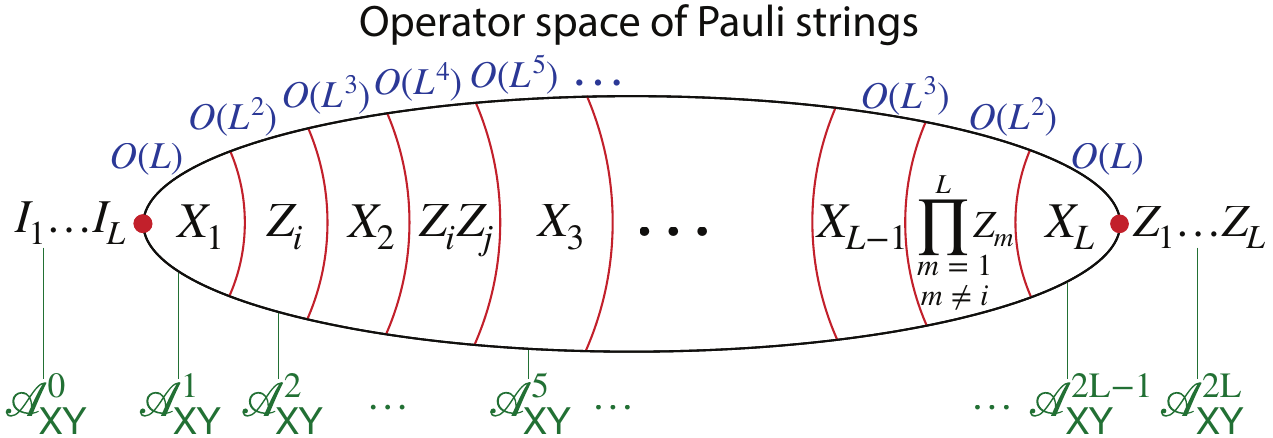}
  \caption{
  Schematic representation of the partitioning of the operator space of Pauli strings into $2L+1$ equivalence classes by the $XY$ Hamiltonian (\ref{hamXY}). Each segment is labeled with a representative observable from the corresponding class. Operators that belong to a certain class never leave it during the evolution. Pauli strings that are integrals of motion are depicted as dots at each end of the ellipse. 
  \label{pic:split}}
\end{figure}

In general, for a given Hamiltonian, the equivalence class of any operator can be determined by algorithmically generating all strings resulting from commutation with the Hamiltonian terms. The algorithm \ref{alg:gen_bas} provides a concrete realization of this procedure.

\begin{algorithm}[t]
  \caption{Generation of equivalence class}\label{alg:gen_bas}
      \begin{algorithmic}
      \Require Fix $A_1$ and $\mathcal{H}=\{H_n\}^N_{n=1}$
      \State $D = 1$
      \State $\text{counter} = 0$
      \While{$\text{counter} < D$}
      \State $V=\mathcal{A}[{\text{counter}+1}]$
      \For{$n=1, \; n\leq N, \; n=n+1$}
      \State $[H_n,V]=a A$
      \If{$a\neq  0$ and $A\notin \mathcal{A}$}
          \State $\mathcal{A}\leftarrow A$
          \Comment{Add $A$ to the subset}
          \State $D=D+1$  
      \EndIf
      \EndFor
      \State $\text{counter}=\text{counter}+1$
      \EndWhile
\end{algorithmic}
\end{algorithm}

Algorithm \ref{alg:gen_bas} adds new Pauli strings obtained from the commutation with strings from $\mathcal{H}$ to the set $\mathcal{A}$ if they are not already in the set. The algorithm stops when no new string can be generated. As a result of this algorithm, we obtain subset $\mathcal{A}$ of dimensionality $D$, which is closed with respect to the commutation with $H$. The efficient implementation of Algorithm \ref{alg:gen_bas} can be achieved by using specialized methods that enable the effective representation of Pauli strings in computer memory and allow for the fast computation of commutation relations between them \cite{uskov2024quantum,dehaene2003clifford}. Recently The Algorithm \ref{alg:gen_bas} has been efficiently implemented in the numerical package PauliStrings.jl \cite{loizeau2024quantum,PauliStringsJL}. 

For a finite $L$, this algorithm must always stop because in this case, the algorithm can generate only $4^L-1$ strings at most. Note   that the identity Pauli string $I^{(1)}=\mathbb{I}\equiv I_1\otimes\dots\otimes  I_L$ is a trivial integral of motion and can never be generated as a commutation of two other strings. In the  simplest case, when $A_1$ is the integral of motion and therefore commutes with the Hamiltonian, no new strings will be added to $\mathcal{A}$, thus $D=1$.

After constructing the equivalence class $\mathcal{A}=\{A_1,\dots,A_D\}$, we can write down the Heisenberg equations in a vectorized form as follows:
\begin{align}
\label{heis_eq_matr}
\frac{d}{dt}A(t)=MA(t),
\end{align}
where we introduced vector $A$ composed of operators as $A=(A_1,\dots,A_D)$, and $A(0),A(t)$ are Schr{\"o}dinger and Heisenberg representations correspondingly. The elements of matrix $M$ are defined as follows:
\begin{align}
\label{matr_M}
m_{ij}=\sum\limits^N_{n=1}ih_n\llangle A_j, \left[ H_n,A_i\right] \rrangle.
\end{align}
It can be easily verified that matrix $M$ is real and skew-symmetric, $m_{ij}=m_{ij}^*=-m_{ji}$. 

To find $A_i(t)$, one must further compute the matrix exponential $S(t)=e^{Mt}$. After this, the solution of (\ref{heis_eq_matr}) can be found as
\begin{align}
\label{pauli_str_heis}
A_i(t)=\sum\limits^D_{j=1}s_{ij}(t)A_j(0),
\end{align}
where $s_{ij}(t)$ are matrix elements of $S(t)$. Notice that computing the matrix exponential of $M(t)$ requires additional computational effort, although polynomial in $D$. However, if one is only interested in the expectation values $\langle A_i\rangle(t)\equiv\tr(A_i(t)\rho_\text{ini})$ for a specific initial state $\rho_\text{ini}$, then direct integration of (\ref{heis_eq_matr}) can be performed, which is a computationally easier task.

In the case of non-integrable dynamics at late times the amplitudes $f_m(t)$ follow a Porter-Thomas distribution, and deviate from it in the integrable case, see Appendix \ref{appendixA}.

\section{Disordered $XY$-spin chains}

Let us consider a simple application of the equivalence class framework by examining the disordered $XY$-spin chain, with Hamiltonian
\begin{align}
\label{hamXY}
H_\text{XY}&=\sum\limits^L_{i=1}J^{xx}_{i}X_iX_{i+1}+J^{yy}_{i}Y_iY_{i+1}+\nonumber\\
&J^{xy}_{i}X_iY_{i+1}+J^{yx}_{i}Y_iX_{i+1}+h^z_iZ_i,
\end{align}
Here the $ J $ and $ h $ coefficients take arbitrary real values.  Unless stated otherwise, we will assume open boundary conditions. 

We start with the well-known classes $\mathcal{A}^1_\text{XY}$ and $\mathcal{A}^2_\text{XY}$, which can be generated from the strings $X_1$ and $Z_i$, respectively. The class $\mathcal{A}^1_\text{XY}$ is known as Majorana strings and has the following elements:
\begin{align}
\label{set_PX1}
\mathcal{A}^1_\text{XY}=&\left\{X_1, Z_1X_2, Z_1Z_2X_3, \dots, Z_1\dots Z_{L-1}X_L, \right. \nonumber \\
&\left. \:\; Y_1, \, Z_1Y_2, \, Z_1Z_2Y_3, \; \dots, Z_1\dots Z_{L-1}Y_L \right\},
\end{align}
which has dimension $D^1_\text{XY}=2L$. The class $\mathcal{A}^2_\text{XY}$ consists of special Pauli strings known as
Onsager strings \cite{Jha_1973,Teretenkov_2024}. An Onsager string is either a single Pauli matrix $Z_j$ or  a product of  Pauli matrices on consecutive sites  with matrices $X$ or $Y$ at the ends and  matrices  $Z$ in the middle, e.g. $Y_3Z_4Z_5Z_6X_7$. There are  $D^2_\text{XY}=2L^2-L$ Onsager strings.  Both these classes have an OED that are low-order polynomials in $ L $.  This can be considered the fundamental reason that particular observables in these classes can be simulated efficiently.  

Interestingly, these are not the only classes that are present for the $ XY $ Hamiltonian.  In fact we find that $\mathcal{P}$ is divided into $K=2L+1$ equivalence classes as $\mathcal{P}=\mathcal{A}^0_\text{XY}\bigcup\dots \bigcup\mathcal{A}^{2L}_\text{XY}$.  The full set of classes can be denoted by $\mathcal{A}^{2n-1}_\text{XY}\equiv\mathcal{A}_\text{XY}[X_n]$ and $\mathcal{A}^{2n}_\text{XY}\equiv\mathcal{A}_\text{XY}[\prod^n_{m=1}Z_m]$, where $n\in [1,L] $.  In addition to these there is also the trivial identity class $\mathcal{A}^0_\text{XY}=\{I_1\otimes\dots \otimes I_L\}$ which is always present in any model. The partition of $\mathcal{P}$ into equivalence classes induced by the Hamiltonian (\ref{hamXY}) is graphically represented in Fig. \ref{pic:split}.

\begin{figure}
\includegraphics[width=\columnwidth]{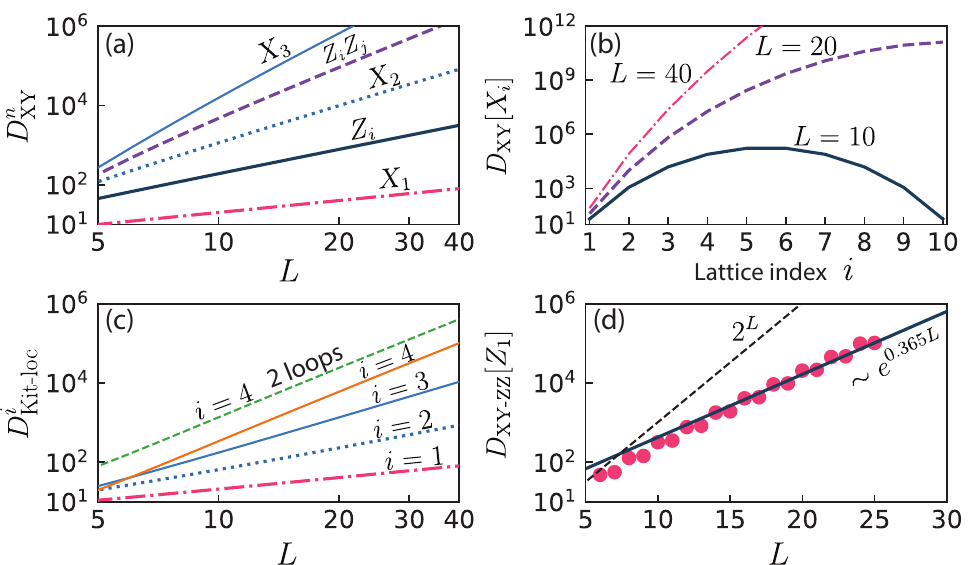}
\caption{
The growth of OEDs for various operators as a function of the system size $L$ and lattice index $i$. Log-log plot illustrating the (a) OEDs of different operators as labeled in the $XY$ chain (\ref{hamXY}); (b) OEDs of operators $X_i$ in the $XY$ chain $\mathcal{A}_\text{XY}[X_i]$ as a function of lattice index $i $; (c) sum of OEDs for single site operators in the Kitaev chain $D^i_\text{Kit-loc}=D_\text{Kit}[X_i]+D_\text{Kit}[Y_i]+D_\text{Kit}[Z_i]$. (d) Semi-log plot showing numerically obtained values of OEDs for $Z_1$ operator in the $XY$-$ZZ$ chain and the corresponding exponential fit $\sim 11.03  e^{0.365413 L}$.
\label{pic:growth}}
\end{figure}

Let us now determine the dimension of these equivalence classes (see also Appendix \ref{appendixB}).  For an arbitrary   system size $L$, the OED of the equivalence class $ \mathcal{A}^{N}_\text{XY}$ can be written as  a polynomial of degree $N$:
\begin{align}
\label{dim_general}
D^N_\text{XY}(L)=\sum\limits^{N}_{j=0}k^N_j L^j.
\end{align}
The coefficients $k^N_0,\dots,k^N_N$ for $N\le L$ are determined as solutions of the following system of equations:

\begin{align}
\label{eqK}
V(0,\dots,N)\vec{K}^N=\vec{D}^N_\text{XY}.
\end{align}
Here $V(0,1,\dots,N)$ is a Vandermonde matrix, $\vec{K}^N\equiv(k^N_0,\dots,k^N_N)$, and the vector of values $\vec{D}^N_\text{XY}=(D^N_\text{XY}(0),\dots,D^N_\text{XY}(N))$ is given by
\begin{align}
\label{vec_vals}
D^N_\text{XY}(m)= \begin{cases} 
0 & \; m \leq \lceil N/2 - 1 \rceil  \\
D^{\lceil 2(m-N/2)\rceil}_\text{XY}(m) & \; m \geq \lceil N/2 \rceil \\
4^N-2\sum\limits^{N-1}_{n=0}D^n_\text{XY}(m) & \; m=N
\end{cases} .
\end{align}
For a specific $N$, the vector of values is determined from the previous $N-1$ polynomials. Specifically, $D^N_\text{XY}(m)$ is equal to zero for the first several values of $m$, as on a lattice of length $m$ there is simply no equivalence class that corresponds to $N$. The remaining values are determined from the relation $D^n_\text{XY}=D^{2L-n}_\text{XY}$. The same relation provides the polynomial in the case $N\ge L+1$. The last line in \eqref{vec_vals} is derived from the fact that the sum of dimensions of all classes must be equal to $4^L$. Therefore, starting from classes $\mathcal{A}^1_\text{XY}$ and $\mathcal{A}^2_\text{XY}$ one can iteratively construct all the other classes and compute their dimensionalities.   Despite the extensive research history of some of these models, to our knowledge the complexity of arbitrary local operators has not been discussed.

For example, the expression for $D^5_\text{XY}$ is found to be: 
\begin{align}
\label{systeqEven}
&D^5_\text{XY}=\frac{4}{10}L-\frac{5}{3}L^2+\frac{7}{3}L^3-\frac{4}{3}L^4+\frac{4}{15}L^5.
\end{align}
The above integer-valued polynomial gives the OED of the operators $X_3$ and $Y_3$. This means that determining the exact dynamics of these operators in the basis of Pauli strings would require solving $D^5_\text{XY}$ linear equations. Figure \ref{pic:growth}(a) shows some other examples on a logarithmic plot, showing the polynomial growth of these equivalence classes.  For an arbitrary lattice index $X_n$, the dimensionality of the corresponding class is given by a polynomial of degree $2n-1$. In a computational complexity sense, this makes calculating the dynamics of any arbitrary Pauli string in the $ XY $ chain efficient, in the sense that the OED is polynomial.  However, towards the middle of the chain the degree of the polynomial may be prohibitively large such that practically larger chains are inaccessible.  In Fig. \ref{pic:growth}(b), we plot the dependence of $D_\text{XY}[X_i]$ on the lattice index $i $, showing a peak in the middle of the chain and smaller values near the edges. This example demonstrates the edge-to-bulk growth of OED within the $XY$-chain. This implies the greater computational overhead to study dynamics of observables in the bulk of the system.  
\begin{figure}
\includegraphics[width=\columnwidth]{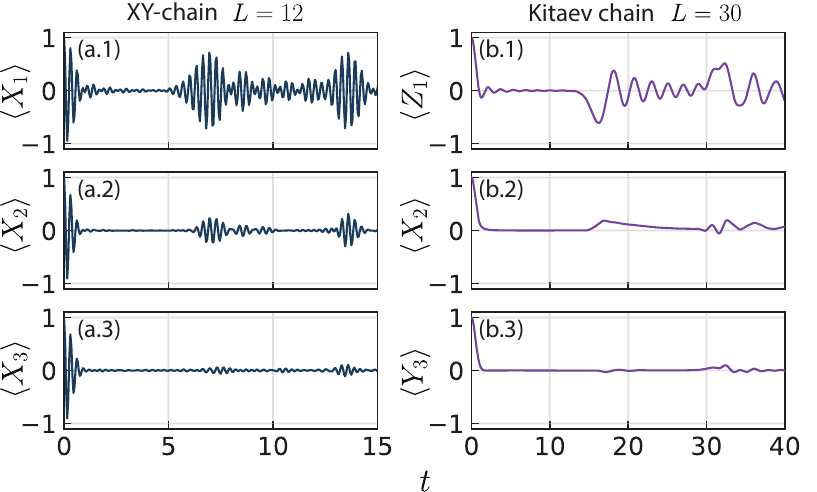}
\caption{
Single-qubit relaxation dynamics for the initial state $\rho_\text{init}=|\psi_i\rangle\langle\psi_i|\otimes I^{\otimes L -1}/2^{L-1} $, here $|\psi_i\rangle$ is the state of $i$-th qubit, the rest of the system is at the infinite temperature. (a) $XY$-chain (\ref{hamXY}), $J^{\alpha\beta}_i=1$, $h^z_i=10$, $L=12$, $|\psi_i\rangle=|+\rangle_i$ (b) 1D Kitaev chain (\ref{ham_kit}), $J^x_i=J^y_i=J^z_i=1$, $L=30$, $|\psi_1\rangle=|1\rangle_1$, $|\psi_2\rangle=|+\rangle_2$, $|\psi_3\rangle=|-\rangle_3$.
\label{pic:edgeBulk}}
\end{figure}

To show that the increase in complexity from the edge to the bulk has also physical ramifications, we compute the relaxation dynamics of a single qubit at infinite temperature.  The $XY$ chain is known to have boundary modes that are fully decoupled from the bulk excitations \cite{kitaev2001unpaired,mi2022noise}.  As such we expect qubits on the edges to maintain coherence to a better extent that those in the bulk.  We consider the $ XY $ Hamiltonian with the uniform couplings $J^{\alpha\beta}_i=1$ and a large magnetic field $h^z_i=10$, to induce fast oscillations in the $XY$-plane so that we may observe the coherence decay \cite{ma2019operational,radhakrishnan2019basis}.  As the initial state, we consider $\rho_\text{init} \propto |+\rangle_i\langle+|_i\otimes I^{\otimes L -1}$, that corresponds to $i$th qubit being fully polarized along $x$-axis and the rest of the system being at the infinite temperature state. Although the system under consideration is integrable, a single qubit placed inside an infinite temperature bath still rapidly relaxes to its equilibrium value. This, however, does not occur for the edge qubit, as its evolution takes place in a small subspace that is decoupled from the bulk. This is manifested by the collapse and revival it undergoes during evolution, as seen in Fig. \ref{pic:edgeBulk}(a.1). We observe that qubits adjacent to the edge qubit also exhibit collapse and revival, although they are less pronounced, with the amplitude vanishing as one moves towards the bulk, see Fig. \ref{pic:edgeBulk}(a.2)(a.3). The revivals can be explained by the fact that the OED of the corresponding operator near the edge is still relatively small, so the initial qubit polarization cannot dissipate swiftly due to the limited number of degrees of freedom available for dissipation.

\section{Mixing of Equivalence Classes by Perturbations and Quenches} 

Introducing perturbations to a Hamiltonian can either preserve or alter its classification of Pauli strings into equivalence classes.

\subsection{Equivalence-Preserving Quenches} 

First, let us consider the case where a quench does not affect the separation of the Hamiltonian into equivalence classes. One of the most interesting scenarios arises in time-dependent Hamiltonians. While our formalism was initially developed under the assumption of a time-independent Hamiltonian for simplicity, the same approach remains applicable to time-dependent Hamiltonians with no additional computational cost.  

Firstly, it is important to note that if a Pauli string $A_n$ belongs to an equivalence class $A_n \in \mathcal{A}$ derived from the set $\mathcal{H}$, a quench of Hamiltonian parameters from $h_n \rightarrow h'_n$ during evolution does not cause $A_n(t)$ to exit $\mathcal{A}$, provided that all $h_n\neq 0$ and no additional strings were added to $\mathcal{H}$.

Indeed, let us consider two evolution operators, $U_1 = e^{-iH\tau_1}$ and $U_2 = e^{-iH'\tau_2}$, where $H$ and $H'$ correspond to the same set of Hamiltonian strings $\mathcal{H}$ but differ in their coefficients, with $h_n \neq h'_n$. The outcome of the algorithm \ref{alg:gen_bas} only depends on the set of Hamiltonian strings $\mathcal{H}=\{H_n\}^N_{n=1}$ not on the coefficients in front of them, therefore the dimensions of the matrix (\ref{matr_M}) does not change after such parameter quench, cf. \cite{gritsev2017integrable}. After sequential applications of $U_1$ and $U_2$, the string $A_n$ transforms as follows:
\begin{align}
\label{quench_parameters}
U^\dagger_2U^\dagger_1A_nU_1U_2=\sum\limits^D_{m=1}\sum\limits^D_{k=1}s^1_{nm}(\tau_1)s^2_{mk}(\tau_2)A_k(0),
\end{align}
here, $s^1_{nm}(\tau_1)$ and $s^2_{mk}(\tau_2)$ are elements of the matrix exponents $S^1(\tau_1)$ and $S^2(\tau_2)$ defined from the matrix (\ref{matr_M}), and $D$ is the OED of the operator $A_n$. The resulting evolution matrix is simply the product of two matrix exponentials $S^{12} = S^1(\tau_1)S^2(\tau_2)$.

In case of time-dependent Hamiltonian, let us assume that the coefficients $h_n$ are now functions of time, $h_n = h_n(t)$, making the matrix (\ref{matr_M}) time-dependent as $M(t)$. Since parameter quenches $h_n \rightarrow h'_n$ do not affect the equivalence class, the matrix (\ref{matr_M}) retains its dimensionality. 

Another example of equivalence-preserving quenches is the application of certain quantum gates during evolution. The simplest case involves Pauli gates $W=X_i,Y_i,Z_i$ which act on any Pauli string as $W^\dagger P_nW=\pm P_n$, leaving any equivalence class of Pauli strings unaffected. A more specific, system-dependent example arises in the equivalence class $\mathcal{A}^2_\text{XY}$, it is easy to check that the phase gate $W=\sqrt{Z_i}$ and the gate $W=X_i\cos\alpha+Y_i\sin\alpha$, where $\alpha\in\mathbb{R}$, are also equivalence preserving quenches. Therefore $XY$ dynamics can be augmented with an arbitrary number of the above quenches without complicating its simulability. This example illustrates non-maximality of $XY$ dynamics with respect to addition of parity-violating unitaries\cite{brod2011extending}.

\subsection{Non-Equivalence-Preserving Quenches}

Now, let us examine how introducing new Hamiltonian strings to $\mathcal{H}$ or applying specific gates can mix different equivalence classes.

First, consider the effect of imposing periodic boundary conditions on the Hamiltonian (\ref{hamXY}). This modification alters the equivalence classes: instead of having a unique class for each operator $X_n$, only two classes distinguished by odd and even lattice indices are present, each exhibiting exponential dimensionality $4^{L-1}$. The class $\mathcal{A}^2_\text{XY}$ remains composed of Onsager strings, but since the lattice is periodic its dimensionality doubles, resulting in $D^2_\text{XY-PBC}=2D^2_\text{XY}$. This example illustrates how the same perturbation can affect OEDs of different operators in a drastically different way.

\begin{figure}
\includegraphics[width=\columnwidth]{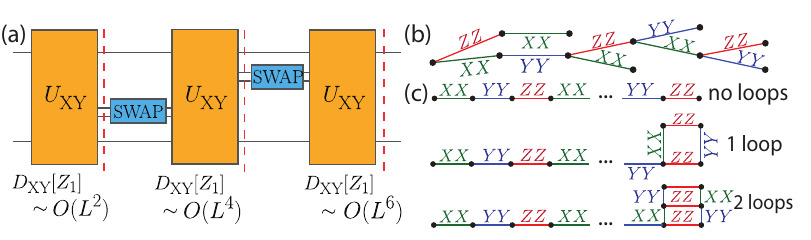}
\caption{
(a) Growth of OED $D_\text{XY}[Z_1]$ of the $Z_1$ operator during the SWAP gate perturbed $XY$-evolution. Blocks $U_\text{XY}=e^{-itH_{XY}}$ describe evolution governed by the Hamiltonian (\ref{hamXY}). As nearest-neighbour SWAP gates are applied, $D_\text{XY}[Z_1]$ increases as $O(L^{2+2m})$, where $m$ is the number of applied gates. (b) Illustration of a Kitaev Hamiltonian (\ref{ham_kit}) on the tree. (c) Geometry of 1D Kitaev chains with and without loops.
\label{pic:pertgraph}}
\end{figure}

Another example of a perturbation is applying $\text{SWAP}$-quenches, when a $ XY $ Hamiltonian evolution is interrupted by a nearest-neighbour $\text{SWAP}$-gate, see Fig. \ref{pic:pertgraph}(a).  Examining the equivalence classes of $Z_i$, it is found that a single $\text{SWAP}_{j,j+1}$-gate application merges classes $\mathcal{A}^2_\text{XY}$ and $\mathcal{A}^4_\text{XY}$, into a unified class with the dimension being fourth order in $L$. Generally, applying multiple $\text{SWAP}$-quenches during $XY$ evolution increases the OED of $Z_i$ as polynomial of degree $2+2m$, with $m$ representing the number of quenches. This is consistent with the asymptotic results recently reported in \cite{mocherla2023extending}.  Ultimately, $\text{SWAP}$-quenches will intermix all equivalence classes into one, achieving a dimensionality of $4^L-1$. This is consistent with the fact that $XY$ dynamics combined with $\text{SWAP}$-quenches is universal \cite{jozsa2008matchgates}. 
%Note that these gates are not parity-preserving which is expected as any parity-preserving non-matchgate unitary makes $XY$ dynamics universal \cite{brod2011extending}.  

\subsection{Other Hamiltonians.} 
To show that our methods are applicable in a more general context, we consider other Hamiltonians. One example is the Kitaev Hamiltonian:  

%on a tree graph  $G^L_\text{tree}$ with $L$ vertexes and at most three edges incident to each vertex (Fig. \ref{pic:pertgraph}(b)). The Hamiltonian has the form:
%
\begin{align}
\label{ham_kit}
H_\text{Kit}=\sum\limits_{\langle i,j\rangle\in G}J^{x}_{ij}X_iX_j+J^{y}_{ij}Y_iY_j+J^{z}_{ij}Z_iZ_j,
\end{align}
where $G$ is a tree graph, and for a given pair $\langle i,j\rangle$ of neighbouring vertexes two of three coupling constants  $J^{x,y,z}_{ij}$  must be zero, but the remaining is non-zero. In addition, the Hamiltonian must have a proper 3-edge coloring of the graph, where each ``color'' represents the interaction type $XX,YY$ or $ZZ$, and proper coloring implies that no vertex belongs to two edges with the same color see Fig. \ref{pic:pertgraph}(b).

Kitaev introduced the Hamiltonian (\ref{ham_kit}) for the hexagonal lattice and showed how to map it to a fermionic model that becomes quadratic after fixing the values of conserved charges \cite{Kitaev}. Later it was shown that the same integrability structure emerges for a large class of tricoordinate lattices, including tree graphs described above \cite{Si_2008,Mandal_2009,Hermanns_2014,OBrien_2016}. Hamiltonians such as (\ref{ham_kit}) offer a rich playground for exploring  equivalence classes and the dynamics of operator growth since they contain numerous local operators that belong to equivalence classes of relatively low dimensions. 

For simplicity let us consider 1D version of the (\ref{ham_kit}). In this case, each local operator $X_n$, $Y_n$ or $Z_n$ belongs to its unique equivalence class and has OEDs that can be given as at most $n$th degree polynomial of $L$ see Fig. \ref{pic:growth}(c). Introducing external magnetic fields or loops in the chain one can merge some of these classes in a controllable manner. For example, in Fig. \ref{pic:growth}(c) we demonstrate how the total OEDs of local operators $X_4,Y_4,Z_4$ increases when the 1D structure is disrupted by adding loops while maintaining its length (Fig. \ref{pic:pertgraph}(c)). This results in a dimensional jump of $D^n_\text{Kit-loc}$ by a factor of $2^mD^n_\text{Kit-loc}$, where $m$ denotes the number of added loops. This relation holds within range of $n$ and $m$, as long as $2^mD^n_\text{Kit-loc}\ll 4^L$.

%We find that OEDs of local operators in the Kitaev chain can always be expressed as an integer-valued polynomial as long as the tree structure of the Hamiltonian is maintained, with the specific geometry influencing the form of the polynomial.

Finally, we introduce the $XY-ZZ$ model which again induces a nontrivial division of equivalence classes.  The Hamiltonian of $XY-ZZ$ model reads:
\begin{align}
\label{xyzz}
H_\text{XY-ZZ}=\sum\limits^{L}_{i=1}J^x_{2i-1}&X_{2i-1}X_{2i}+J^y_{2i-1}Y_{2i-1}Y_{2i}\nonumber\\
&+J^z_{2i}Z_{2i}Z_{2i+1} ,
\end{align}
here again the coefficients $J$ take arbitrary real values. In case of this Hamiltonian the operators $X_i$ and $Y_i$ each belong to distinct equivalence classes with dimensionalities $D_\text{XY-ZZ}[X_i,Y_i]=4^{L-3}$. Pairs of operators $Z_{2i-1}$ and $Z_{2i}$ for $i>1$, belong to equivalence classes with slightly bigger dimensionality scaling as $O(4^{L-3})$.  While the OED of these classes are exponentially growing, some grow at a much slower rate $2^L$. The most intriguing case involves the edge operator $Z_1$, whose equivalence class dimensionality grows approximately as $D_\text{XY-ZZ}[Z_1]\simeq 11 e^{0.365L}$, see Fig \ref{pic:growth}(d). The factor $0.365$ in the exponent enables a significant practical reduction in problem complexity, compared to exact diagonalization. 

%For example, if one were to calculate the dynamics of $Z_1(t)$ using exact diagonalization, even for a fixed initial state, it would require $O(2^L)$ resources. Suppose that one has enough computational power to perform exact diagonalization for $30$ qubits, then by solving Heisenberg equations inside of the equivalence class $\mathcal{A}^{Z_1}_\text{XY-ZZ}$ one could find $Z_1(t)$ for up to $45$ qubits with the same resources.

\subsection{Dissipative dynamics.} Finally, we also mention that the partitioning of the operator space to equivalence classes also takes place for dissipative dynamics that is described by GKSL equation \cite{breuer2002theory}. Furthermore, in certain cases, dissipation also does not violate the structure of certain equivalence classes or violates it insignificantly. This is the case e.g. for the Hamiltonian (\ref{hamXY}) with dephasing \cite{vznidarivc2010exact,vznidarivc2013transport,vznidarivc2010exact,shibata2019dissipative,shibata2019dissipativeKit,Dolgirev_2020,ghosh2023relaxation,chen2024many}. Thus, our methods can be directly applied to treat various (semi-)integrable dissipative models \cite{Medvedyeva_2016,ermakov2024effect,Teretenkov_2024,Ferreira_2024} as well as for finding new ones. 

Indeed, let us consider the scenario where the dynamic of a certain operator $A_1$ is governed by the GKSL equation \cite{breuer2002theory}, that reads:
\begin{align}
\label{gksl_eq_sup}
\frac{d}{dt}A_1=i[H,A_1]+\mathcal{D}^\dagger A_1,
\end{align}
where $\mathcal{D}^\dagger$ is an adjoint dissipation superoperator defined as:
\begin{align}
\label{d_op_sup}
&\mathcal{D}^\dagger=\sum^M_{m=1}\gamma_m\mathcal{D}^\dagger_m,\nonumber\\
&\mathcal{D}^\dagger_mA_1=l^\dagger_mA_1 l_m-\frac{1}{2}\{l^\dagger_m l_m,A_1\}
\end{align}
here $l_m$ are Lindblad operators, $\gamma_m$ are real positive constants, and $\{\cdot,\cdot\}$ denotes the anticommutator. We assume that all the Lindblad operators are Pauli strings.

\begin{algorithm}[t]
\caption{Generation of equivalence class in dissipative case}\label{alg:gen_bas_diss}
\begin{algorithmic}
\Require Fix $A_1$, $\mathcal{H}=\{H_n\}^N_{n=1}$ and $\{\mathcal{D}^\dagger_m\}^M_{m=1}$
\State $D = 1$
\State $\text{counter} = 0$
\While{$\text{counter} < D$}
\State $V=\mathcal{A}[{\text{counter}+1}]$
\For{$n=1, \; n\leq N+M, \; n=n+1$}
\If{$n\leq  N$}
    \State $[H_n,V]=a A$
\Else
    \State $\mathcal{D}^\dagger_{n-N}A=a A$
\EndIf
\If{$a\neq  0$ and $A\notin \mathcal{A}$}
    \State $\mathcal{A}\leftarrow A$
    \Comment{Add $A$ to the subset}
    \State $D=D+1$  
\EndIf
\EndFor
\State $\text{counter}=\text{counter}+1$
\EndWhile
\end{algorithmic}
\end{algorithm}

New Pauli string can be generated from an existing one by commutation with a Hamiltonian string $[H_n,\cdot]$, and as a result of action of dissipation superoperator $\mathcal{D}^\dagger_m$. Consequently, algorithm \ref{alg:gen_bas} must be modified. The modified version is presented as algorithm \ref{alg:gen_bas_diss}. Additionally, the matrix (\ref{matr_M}) needs to be updated as follows:
\begin{align}
\label{matr_M_diss}
m_{ij}=\sum\limits^N_{n=1}ih_n\llangle A_j, \left[ H_n,A_i\right] \rrangle+\sum\limits^M_{m=1}\gamma_m\llangle A_j, \mathcal{D}^\dagger_mA_i \rrangle.
\end{align}
After computing the matrix $M$, one can use either formula (\ref{heis_eq_matr}) or (\ref{pauli_str_heis}) to determine the dynamics of the observable of interest.

\subsection{Discussion.} We have developed a formalism of equivalence classes which allows one to obtain the dimension of the operator evolution in the Heisenberg picture for spin-$1/2$ Hamiltonians. The approach gives a straightforward way of understanding the nature of why particular models can be time integrable, and provides a powerful tool for identifying when quantum dynamics are simulable. We examined well-studied models such as the $XY$ and Kitaev chains and recovered the known result that all $ Z_i$ can be efficiently computed. Despite the long history of these models, to our knowledge, the remaining equivalence classes have not been identified to date. We showed that the $X_i, Y_i$ observables increase in complexity moving from edge to bulk. We also considered various perturbations that may or may not affect the OEDs of certain operators. We showed that applications of certain single-qubit gates during the evolution do not affect OED of some operators under simulation. 

%This is interesting from a quantum circuit point of view as it implies that matchgate circuits can be augmented with such gates without penalty regarding the simulation. We examined universality-enabling gates for $XY$ evolution, specifically SWAP gates, which increases the degree of the OED linearly with the number of SWAP gates.  Similarly, loops on Kitaev model on trees increases the complexity by an exponential factor in the number of loops.   

%To illustrate our results we mainly showed results on free-fermionic models, but they may also be applied in other contexts. 

%, showing that all equivalence classes have exponential dimensionality

We introduced the non-free fermion $XY$-$ZZ$ model. In this model, almost each local operator still resides within its own equivalence class. Moreover, the $z$-projections of edge qubits also have OEDs that scale more slowly than those of bulk qubits. Solving the dynamics within the corresponding equivalence class will still significantly outperform brute force approach that disregard the splitting of the operator space, such as exact diagonalization. While we focused on the basis of Pauli strings for simplicity in this paper, it is important to note that operators with small OED exist in other bases as well \cite{uglov1996sl,miao2022generalised,gamayun2022out,lychkovskiy2021closed}. Therefore, studying the separation of other bases into equivalence classes may potentially yield more examples of simulable quantum dynamics.

\noindent{\it Note added}. An approach conceptually similar to ours has been recently proposed in \cite{grigoletto2024exact}.

\begin{acknowledgments}
I. E. and O. L. are supported by Rosatom in the framework of the Roadmap for Quantum computing (Contract No. 868-1.3-15/15-2021 dated October 5,2021). Part of the work presented in Section III ``Disordered XY-spin chains'' was supported by the Foundation for the Advancement of Theoretical Physics and Mathematics “BASIS” under the grant N 22-1-2-55-1. T. B. is supported by the SMEC Scientific Research Innovation Project (2023ZKZD55); the Science and Technology Commission of Shanghai Municipality (22ZR1444600); the NYU Shanghai Boost Fund; the China Foreign Experts Program (G2021013002L); the NYU-ECNU Institute of Physics at NYU Shanghai; the NYU Shanghai Major-Grants Seed Fund; and Tamkeen under the NYU Abu Dhabi Research Institute grant CG008.
\end{acknowledgments}

\appendix
\section{Amplitudes in Heisenberg representation}
\label{appendixA}

In this Appendix, we apply Algorithm \ref{alg:gen_bas} to various Hamiltonians and operators to analyze the corresponding amplitudes in the Heisenberg representation.

Consider the following Hamiltonians:
\begin{align}
\label{hams1}
&H_\text{1}=-\sum\limits^{L-1}_{i=1}\left(X_iX_{i+1}+2Y_iY_{i+1}\right)+\sum\limits^{L}_{i=1}\left(X_i+Y_i\right),\\
\label{hams2}
&H_\text{2}=\sum\limits^{L-1}_{i=1}\left(X_iX_{i+1}+Y_iY_{i+1}+X_iY_{i+1}+Y_iX_{i+1}\right)-\sum\limits^{L}_{i=1}Z_i,\\
\label{hams3}
&H_\text{3}=\sum\limits^{L-1}_{i=1}\left(X_{3i-2}X_{3i-1}+Y_{3i-1}Y_{3i}+Z_{3i}Z_{3i+1}\right),\\
\label{hams4}
&H_\text{4}=\sum\limits^{L-1}_{i=1}X_{2i-1}X_{2i}+Y_{2i-1}Y_{2i}+Z_{2i}Z_{2i+1}
\end{align}
here, $H_2$, $H_3$, and $H_4$ are specific realizations of the Hamiltonians $H_\text{XY}$, $H_\text{Kit}$, and $H_\text{XY-ZZ}$, respectively, that were introduced earlier in the paper. The Hamiltonian $H_1$ serves as an example of a robustly non-integrable Hamiltonian, as evidenced by its level-spacing statistics \cite{atas2013distribution}.

To compute amplitudes in the Heisenberg representation at a specific time, one should first fix the seed operator $A_1$ and the Hamiltonian as prescribed by Algorithm \ref{alg:gen_bas}. After constructing the corresponding equivalence class for the selected observable, one needs to compute the matrix (\ref{heis_eq_matr}) and the corresponding matrix exponential for the specified time $t$. Amplitudes are given as the square roots of the corresponding line of the matrix exponential $s_{1j}(t)$. We consider the ordered probability amplitudes $[s^2_{1j}(t)]_{P(j)}$, where $P(j)$ is a permutation function that reorders the values from smallest to largest.

\begin{figure*}
\includegraphics[width=2.0\columnwidth]{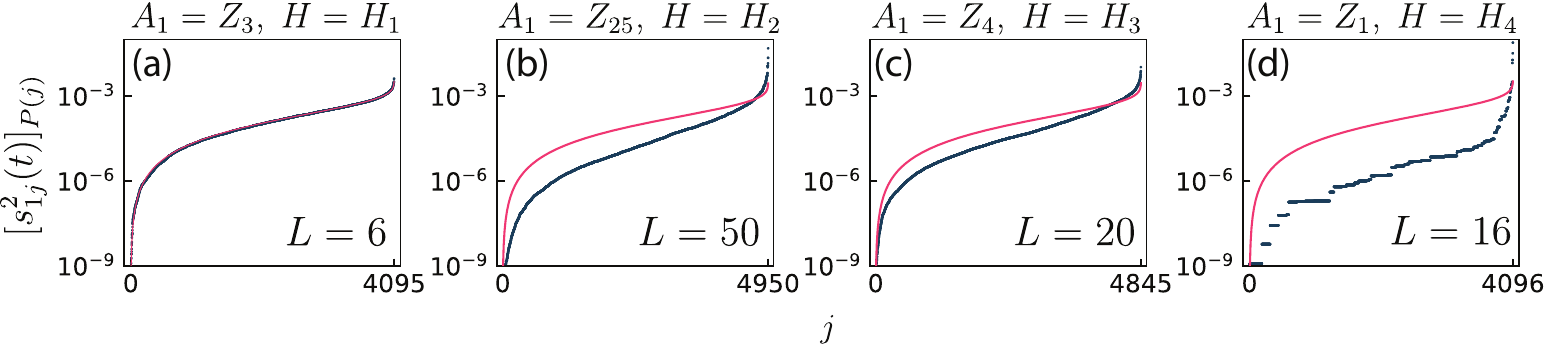}
\caption{Semi-log plot of the distribution of amplitudes $s^2_{1j}(t)$ for various Hamiltonians (\ref{hams1},\ref{hams2},\ref{hams3},\ref{hams4}) and seed operators at time $t=10$. The algorithm \ref{alg:gen_bas} is applied to generate the corresponding equivalence class, after which the coefficients $s^2_{1j}(t)$ are computed and sorted from smallest to largest. The solid pink line represents the Porter-Thomas distribution (\ref{pt_distr}).
\label{pic:s_probdistr}}
\end{figure*}

In Fig. \ref{pic:s_probdistr}, we plot the ordered amplitudes $[s^2_{1j}(t)]_{P(j)}$ for Hamiltonians (\ref{hams1}, \ref{hams2}, \ref{hams3}, \ref{hams4}) and different seed operators $A_1$, fixing time $t=10$ for all cases. In each example, the OED of operator under consideration varies. We also fit our numerical data with the Porter-Thomas distribution \cite{porter1956fluctuations,boixo2018characterizing}, adapted for the Heisenberg representation in \cite{ermakov2024unified}. The formula for this distribution reads:
\begin{align}
\label{pt_distr}
\text{PT}_j[A_1]=\frac{2}{D[A_1]}\left(\text{erf}^{-1}\frac{j}{D[A_1]}\right)^2,
\end{align}
where $A_1$ is the seed operator under consideration and $D[A_1]$ is its OED.

For Hamiltonian (\ref{hams1}), we observe that the OED of operator $Z_3$ is maximal, $D_{H_1}[Z_3]=4^L-1$, which is expected as Hamiltonian (\ref{hams1}) is far from integrability. Furthermore, only in this case do the amplitudes distribution obey the Porter-Thomas distribution (\ref{pt_distr}), which is expected since this distribution characterizes universal quantum evolution.

For other Hamiltonians (\ref{hams2}, \ref{hams3}, \ref{hams4}), we observe that the distribution of amplitudes does not follow the Porter-Thomas distribution. This is also expected, as none of these Hamiltonians can generate universal quantum evolution. Notably, while free-fermionic Hamiltonians (\ref{hams2}, \ref{hams3}) have distributions that appear to be similar, Hamiltonian (\ref{hams4}) exhibits a distinctly different distribution, highlighting its non-free-fermionic nature.

\begin{figure*}
\includegraphics[width=2.0\columnwidth]{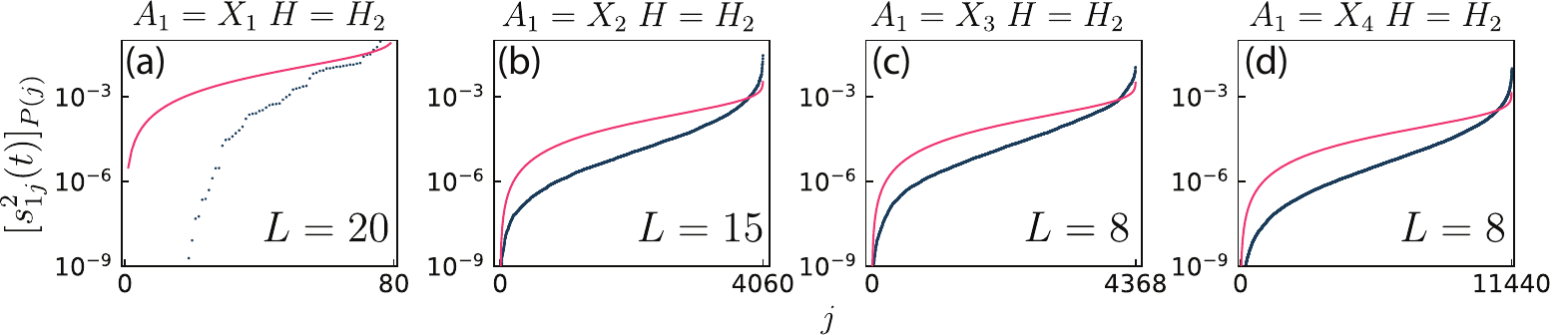}
\caption{Semi-log plot of the distribution of amplitudes $s^2_{1j}(t)$ for various Hamiltonians the Hamiltonian (\ref{hams2}) and seed operators $X_i$ at time $t=10$. The algorithm \ref{alg:gen_bas} is applied to generate the corresponding equivalence class, after which the coefficients $s^2_{1j}(t)$ are computed and sorted from smallest to largest. The solid pink line represents the Porter-Thomas distribution (\ref{pt_distr}).
\label{pic:s_probdistr_x}}
\end{figure*}

Lastly, let us consider different seed observables $X_i$ for a fixed Hamiltonian (\ref{hams2}). In Fig. \ref{pic:s_probdistr_x}, we see that the distribution of amplitudes in this case also does not follow the Porter-Thomas distribution. Furthermore, distribution of amplitudes for the operator on the edge $X_1$ is notably different from that of bulk operators $X_i$.
 
\section{Dimensionalities of Equivalence Classes in the XY Chain}
\label{appendixB}

In this appendix, we derive integer-valued polynomials that exactly describe the dimensionalities of the equivalence classes of the $XY$ Hamiltonian (\ref{hamXY}) with open boundary conditions.

\subsection{Majorana and Onsager strings}

There is always a trivial class $\mathcal{A}^0_\text{XY}$ which consists of identity operator $I^{(1)}=I_1\dots I_L$, the dimensionality of this class is $D^0_\text{XY}=1$.

Let us start with the class of Majorana strings $\mathcal{A}^1_\text{XY}$ which can be generated from the operator $X_1$. The class $\mathcal{A}^1_\text{XY}$ has the following structure:
\begin{align}
\label{set_PX1}
\mathcal{A}^1_\text{XY}=&\left\{X_1, Z_1X_2, Z_1Z_2X_3, \dots, Z_1\dots Z_{L-1}X_L, \right. \nonumber \\
&\left. \:\; Y_1, \, Z_1Y_2, \, Z_1Z_2Y_3, \; \dots, Z_1\dots Z_{L-1}Y_L \right\},
\end{align}
with the dimensionality of $D^1_\text{XY}=2L$. 

The class of Onsager strings $\mathcal{A}^2_\text{XY}$ has the following structure:

\begin{align}
\label{p_structure}
\mathcal{A}^2_\text{XY}=\left\{ \left\{ R^{n}_j \right\}^{L-n,L-1}_{j=1,n=1-L},  \left\{ Q^{n}_j \right\}^{L-n,L-1}_{j=1,n=1-L,n\neq 0}\right\},
\end{align}

where $R^n_j$ and $Q^n_j$ are defined as

\begin{align}
\label{r_and_q}
&R^0_j=Z_j\nonumber\\
&R^{-n}_j=Y_j\left(\prod\limits^{n-1}_{m=1}Z_{j+m}\right)Y_{j+n},\nonumber\\
&R^{n}_j=X_j\left(\prod\limits^{n-1}_{m=1}Z_{j+m}\right)X_{j+n},\\
&Q^{n}_j=X_j\left(\prod\limits^{n-1}_{m=1}Z_{j+m}\right)Y_{j+n},\nonumber\\
&Q^{-n}_j=Y_j\left(\prod\limits^{n-1}_{m=1}Z_{j+m}\right)X_{j+n}.\nonumber
\end{align}

The dimensionality of this class is given by the polynomial $D^2_\text{XY}=2L^2-L$.

\subsection{Mirrored classes}

Equivalence classes possess particular symmetries that may be exploited to deduce their structure.  We first consider the following lemma.
\begin{lemma}
If a product of two Pauli strings $P_{AB}=P_AP_B$ from equivalence classes $P_A\in\mathcal{A}$ and $P_B\in\mathcal{B}$, does not belong to either of them, $P_{AB}\notin\mathcal{A}\cup\mathcal{B}$, then it belongs to an equivalence class $\mathcal{AB}$ which consist only of products of strings from $\mathcal{A}$ and $\mathcal{B}$.
\end{lemma}

To prove the above lemma one needs to use the fact that the following is satisfied for three arbitrary Pauli strings $[P_1P_2,P_3]=\alpha [P_1,P_3]P_2$, here $\alpha\in\mathbb{C}$. In this case let us consider some string generated from $P_{AB}$ and some Hamiltonian string $H_n$:
\begin{align}
\label{lemma_proof_1}
[P_{AB},H_n]\equiv[P_AP_B,H_n]=\alpha P_A[P_B,H_n],
\end{align}
which is a product of two Pauli strings from $\mathcal{A}$ and $\mathcal{B}$ or zero.

Now, let us introduce mirrored classes in the $XY$ chain. The Hamiltonian (\ref{hamXY}), in general, has only one integral of motion that can be expressed as a single Pauli string, namely the operator $I^{(2)}=Z_1\dots Z_L$ which constitutes the class $\mathcal{A}^{2L}_\text{XY}$. 
If one multiplies each Pauli string from an equivalence class $\cal A$ by the  integral of motion $I^{(2)}$, one obtains another,``mirrored'' equivalence class $\overline{\cal A}$, thanks to the identity  $[H_n,I^{(2)}A_k]=I^{(2)}[H_n,A_k]$. More importantly the dimensionality of the original and mirrored classes coincide. Therefore in case of $XY$ chain we have: 
\begin{align}
\label{mirrored}
D^n_\text{XY}=D^{2L-n}_\text{XY}, \qquad n\in [0,L]
\end{align}

Also it is important to note that the class $\mathcal{A}^L_\text{XY}$ always coincides with its mirroring. Indeed it can be is either generated from the operator $X_{(L+1)/2}$ in case of odd $L$ and from $Z_1\dots Z_{L/2}$ in case of even $L$. Below we will also show that this class has maximal dimensionality.

\subsection{Integer-valued polynomials for dimensionalities of equivalence classes}

We now derive integer-valued polynomials that describe dimenionalities of equivalence classes $\mathcal{A}^n_\text{XY}$ in the $XY$ chain. 

First let us consider the simplest case of single qubit $L=1$. In this case the partition into equivalence classes given by the Hamiltonian (\ref{hamXY}) is very simple, namely: $\mathcal{A}^0_\text{XY}=\{I_1\}$, $\mathcal{A}^1_\text{XY}=\{X_1,Y_1\}$ and $\mathcal{A}^2_\text{XY}=\{Z_1\}$. Only $L$ first classes are important to us, because other $L$ classes can be obtained from them by mirroring. In case of $L=1$ it is easy to see that $\mathcal{A}^2_\text{XY}$ is the mirror class to $\mathcal{A}^0_\text{XY}$. Therefore it is evident that:
\begin{align}
\label{pol_L1}
4^1=2D^0_\text{XY}(1)+D^1_\text{XY}(1), \qquad L=1   .
\end{align}

Next we consider the case of two qubits $L=2$. In this case we have classes $\mathcal{A}^0_\text{XY}$ and $\mathcal{A}^1_\text{XY}$ as well as their mirrored classes, additionally to it we have $\mathcal{A}^2_\text{XY}$ with dimensionality $D^2_\text{XY}$. One can check that: 
\begin{align}
\label{pol_L2}
4^2=2D^0_\text{XY}(2)+2D^1_\text{XY}(2)+D^2_\text{XY}(2), \qquad L=2
\end{align}

Let us move one to the case of $L=3$. In this case the class $\mathcal{A}^3_\text{XY}$ generated from the operator $X_2$ appears. Continuing previous line of thinking one can write down:
\begin{align}
\label{pol_L3}
4^3=2D^0_\text{XY}(3)+2D^1_\text{XY}(3)+2D^2_\text{XY}(3)+D^3_\text{XY}(3), \qquad L=3
\end{align}
Unlike previous cases $L=1,2$, here we do not know the general expression for $D^3_\text{XY}(L)$. This can be obtained in the following way. 
The number of strings in $\mathcal{A}^3_\text{XY}$ scales as $O(L^3)$ because it must be proportional to the product of $D^1_\text{XY}$ and $D^2_\text{XY}$. Indeed, $X_2=Z_1\otimes Z_1X_2$, therefore due to the above lemma the class $\mathcal{A}^3_\text{XY}$ consists of products of strings from $\mathcal{A}^1_\text{XY}$ and $\mathcal{A}^2_\text{XY}$. So at most $\mathcal{A}^3_\text{XY}$ can have $D^1_\text{XY}D^2_\text{XY}$ Pauli strings. Since many products from $\mathcal{A}^1_\text{XY}$ and $\mathcal{A}^2_\text{XY}$ yield either the same Pauli string or a string from one of these classes $D^3_\text{XY}<D^1_\text{XY}D^2_\text{XY}$. Therefore, at most $D^3_\text{XY}(L)$ can be an integer-valued polynomial of 3rd order. Let us seek for it in the form:
\begin{align}
\label{pol_L3_ans}
D^3_\text{XY}(L)=k^3_0+k^3_1L+k^3_2L^2+k^3_3L^3.
\end{align}
To determine the coefficients $k^3_0,k^3_1,k^3_2,k^3_3$, we need 4 equations. First, $D^3_\text{XY}(0)=0$ and therefore $k^3_0=0$. This is degenerate case which yet must be considered. Second, $D^3_\text{XY}(1)=0$, because no operators $X_2$ appear in the lattice of $L=1$ qubit. Third equation is $D^3_\text{XY}(2)=D^1_\text{XY}(2)$, because when $L=2$ the class $\mathcal{A}^3_\text{XY}$ generated from $X_2$ is the mirrored to the class $\mathcal{A}^1_\text{XY}$. The last equation can be obtained from the expression (\ref{pol_L3}).

We therefore have the following system of equations
\begin{align}
\label{pol_L3_equations}
&D^3_\text{XY}(0)=0\nonumber\\
&D^3_\text{XY}(1)=0\\
&D^3_\text{XY}(2)=D^1_\text{XY}(2)\nonumber\\
&D^3_\text{XY}(3)=4^3-2D^0_\text{XY}(3)-2D^1_\text{XY}(3)-2D^2_\text{XY}(3).\nonumber
\end{align}
This is the system of linear equations with respect to variables $k^3_0,k^3_1,k^3_2,k^3_3$. By solving it we obtain: 
\begin{align}
\label{pol_L3_sol}
D^3_\text{XY}=\frac{2}{3}L-2L^2+\frac{4}{3}L^3.
\end{align}

The above procedure can be generalized for the case of arbitrary $D^N_\text{XY}$. Let us look for the $D^N_\text{XY}$ in the form of the following polynomial:
\begin{align}
\label{pol_LN_ans}
D^N_\text{XY}=\sum\limits^N_{n=0}k^N_nL^n,
\end{align}
the coefficients $k^N_0,\dots,k^N_N$ are found as the solution of the following system of equations, if $N$ is odd:
\begin{align}
\label{systeqOdd}
&D^N_\text{XY}(0)=0 \nonumber \\
&\qquad\qquad\vdots \nonumber \\
&D^N_\text{XY}\left(\frac{N-1}{2}\right)=0 \nonumber \\
&D^N_\text{XY}\left(\frac{N+1}{2}\right)=D^1_\text{XY}\left(\frac{N+1}{2}\right) \nonumber \\
&D^N_\text{XY}\left(\frac{N+1}{2}+1\right)=D^3_\text{XY}\left(\frac{N+1}{2}+1\right) \nonumber \\
&\qquad\qquad\vdots  \\
&D^N_\text{XY}(N-1)=D^{N-2}_\text{XY}(N-1) \nonumber \\
&D^N_\text{XY}(N)=4^N-2\sum\limits^{N-1}_{n=0}D^n_\text{XY}(N) , \nonumber 
\end{align}
and if $N$ is even then the system changes to:
\begin{align}
\label{systeqEven}
&D^N_\text{XY}(0)=0 \nonumber \\
&\qquad\qquad\vdots \nonumber \\
&D^N_\text{XY}\left(\frac{N}{2}-1\right)=0 \nonumber \\
&D^N_\text{XY}\left(\frac{N}{2}\right)=D^0_\text{XY}\left(\frac{N}{2}\right) \nonumber \\
&D^N_\text{XY}\left(\frac{N}{2}+1\right)=D^2_\text{XY}\left(\frac{N}{2}+1\right) \nonumber \\
&\qquad\qquad\vdots  \\
&D^N_\text{XY}(N-1)=D^{N-2}_\text{XY}(N-1) \nonumber \\
&D^N_\text{XY}(N)=4^N-2\sum\limits^{N-1}_{n=0}D^n_\text{XY}(N) . \nonumber 
\end{align}
The left-hand side of the above systems can be expressed in terms of Vandermonde matrix as $V(0,\dots,N)\vec{K}^N$. The Vandermonde matrix $V(0,\dots,N)$ always have a positive determinant, ensuring that a unique solution exists. 

By consequently solving this system for different $N$ starting from 3 we can obtain expressions for first several polynomials:

\begin{align}
\label{systeqEven}
&D^1_\text{XY}=2L\nonumber \\
&D^2_\text{XY}=-L+2L^2\nonumber \\
&D^3_\text{XY}=\frac{2}{3}L-2L^2+\frac{4}{3}L^3 \\
&D^4_\text{XY}=-\frac{1}{2}L+\frac{11}{6}L^2-2L^3+\frac{2}{3}L^4\nonumber \\
&D^5_\text{XY}=\frac{4}{10}L-\frac{5}{3}L^2+\frac{7}{3}L^3-\frac{4}{3}L^4+\frac{4}{15}L^5\nonumber \\
&D^6_\text{XY}=-\frac{1}{3}L+\frac{137}{90}L^2-\frac{5}{2}L^3+\frac{17}{9}L^4-\frac{2}{3}L^5+\frac{4}{45}L^6\nonumber \\
&D^7_\text{XY}=\frac{2}{3}L-\frac{7}{5}L^2+\frac{116}{45}L^3-\frac{7}{3}L^4+\frac{10}{9}L^5-\frac{4}{15}L^6+\frac{8}{315}L^7  . \nonumber 
\end{align}
By continuing this process one can obtain exact expressions for OEDs of any operator in the $XY$ chain. 

\subsection{Scaling of the leading coefficient in $D^L_\text{XY}$}

\begin{figure}[H]
\includegraphics[width=1.\columnwidth]{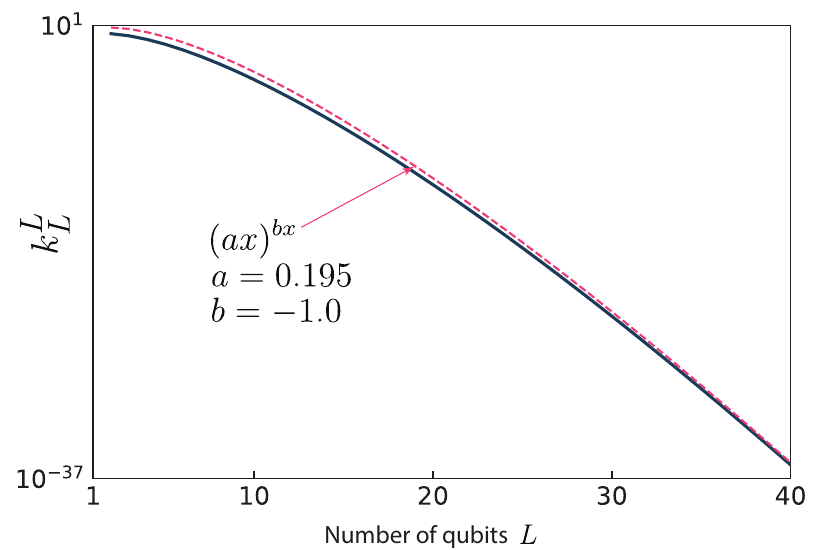}
\caption{The leading coefficient $k^L_L$ in the polynomial (\ref{pol_LN_ans}) corresponding to the equivalence class of maximal dimensionality as a function of $L$. The dashed line represents the superexponential fit $(ax)^{bx}$, where $a=0.195$ and $b=-1.0$.
\label{pic:supCoefAI}}
\end{figure}

The reader might be concerned by the following: for $L$ qubits, there is an equivalence class of maximal dimensionality $D^L_\text{XY}$ scaling as $O(L^L)$. Such superexponential scaling is capable of surpassing the total number of Pauli strings, $4^L$ which would be a contradiction. This, however, never happens as the coefficient $k^L_L$ in front of the leading power in the expression for $D^L_\text{XY}$ also decreases superexponentially; see Fig. \ref{pic:supCoefAI}. 

\bibliographystyle{unsrt}
\bibliography{ref}
%\printbibliography

\end{document}